\documentclass{PoS}

\title{INTEGRAL Results on Gamma-Ray Bursts}

\ShortTitle{INTEGRAL Results on Gamma-Ray Bursts}

\author{\speaker{Diego G\"otz}%
        \thanks{Based on observations with INTEGRAL, an ESA project with instruments and
science data centre funded by ESA member states (especially the PI
countries: Denmark, France, Germany, Italy, Switzerland, Spain), Czech
Republic and Poland, and with the participation of Russia and the USA}\\
       AIM - CEA Saclay DSM/Irfu/SAp, F-91191 Gif-sur-Yvette, France\\
       E-mail: \email{diego.gotz@cea.fr}}

\abstract{Despite being a general observatory, and not a Gamma-Ray Bursts (GRBs) oriented mission, INTEGRAL has contributed to several important discoveries in the GRB field. This has been obtained thanks to its unprecedented localization capabilities, and sensitivity in the soft gamma-ray domain. \\
In this paper I will review the main results obtained during the last 10 years with, and thanks to, INTEGRAL GRBs, including the discovery of one of the few GRBs spectroscopically associated with a Supernova, the first measurement of variable polarization in the GRB prompt emission, the indication of the existence of a low-luminosity population of GRBs, as well as the recent application of GRBs as probes for the fundamental physics. I will mention the main global characteristics of the INTEGRAL sample, and make the point on the lessons learnt from INTEGRAL in the perspective of designing future GRB dedicated missions.}

\FullConference{ An INTEGRAL view of the high-energy sky (the first 10 years) - 9th INTEGRAL Workshop and celebration of the 10th anniversary of the launch\\
		 15-19 October 2012\\
		 Bibliotheque Nationale de France, Paris, France}

\begin{document}

\section{Introduction}
The ESA INTEGRAL mission \cite{integral} has been launched on October 17$^{th}$ 2002, and after ten years of operations it has detected and localized in real-time more than 90 Gamma-Ray Bursts (GRBs; for a recent review see \cite{atteia13}), thanks to the INTEGRAL Burst Alert System (IBAS; \cite{ibas,mereghetti13}). IBAS analyses in real-time the IBIS/ISGRI \cite{ibis,isgri} telemetry reaching the INTEGRAL Science Data Centre (ISDC; \cite{isdc}), and detects and localizes GRBs, and other types of transients appearing in the IBIS field of view during pointed observations.

In this paper I will review the global properties of INTEGRAL GRBs, and some of the most important results that have been obtained through the multi-wavelength follow-up of a few of them.

\section{Global Properties}
\begin{figure}
   \centering
   \includegraphics[width=0.7\textwidth]{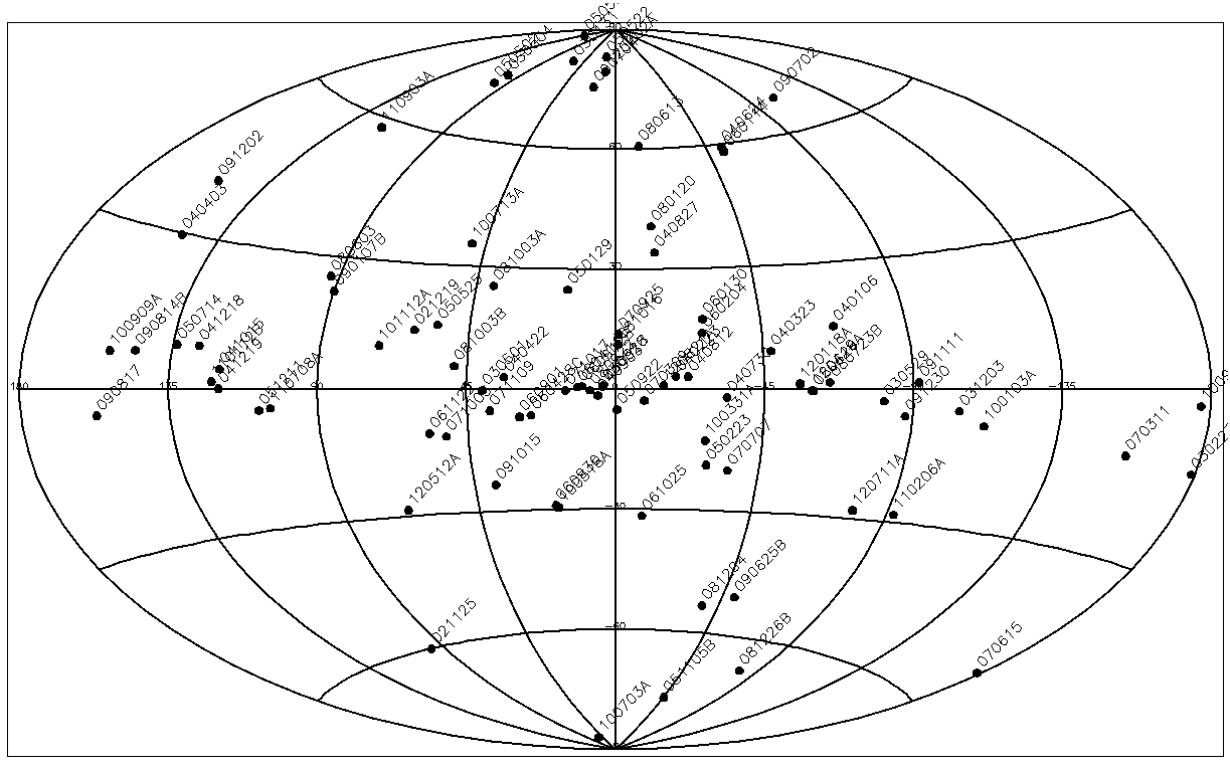}
   \caption{Distribution of INTEGRAL Gamma-Ray Bursts in Galactic Coordinates (Courtesy S. Mereghetti)}
   \label{fig:distr}
   \end{figure}

Out of the 93 GRBs detected so far by INTEGRAL, only 7 have a measured redshift. This is mainly due to the bias in the INTEGRAL pointings that are concentrated towards the Galactic plane, where the high absorption in the optical and near infra-red bands makes the identification of the afterglows of the INTEGRAL bursts particularly difficult, see Fig. \ref{fig:distr}.

In spite of having good multi wavelength coverage, INTEGRAL GRBs can be studied in detail from the spectral point of view, due to the high combined sensitivity of IBIS and SPI \cite{spi} over the 20 keV to 2 MeV energy range.
In order to investigate the spectral properties of the whole INTEGRAL sample,
Bo\v snjak et al. \cite{bosnjak13} have reanalysed all publicly available IBIS/ISGRI and SPI data with an ad-hoc developed software. The combined analysis shows that the INTEGRAL bursts are on average fainter and softer (i.e. they have a lower peak energy) than the ones detected by BATSE, while the distribution of their $T_{90}$ duration is more similar to the one of Fermi/GBM and Swift/BAT bursts than to the one of BATSE, especially due to the lack of short ($T_{90}$ < 2 s) bursts. The latter can be explained, on one hand, by the lower energy threshold of those instruments with respect to BATSE, that makes them more sensible to long-soft events, rather than to short-hard ones, and on the other hand, by the intrinsic paucity of photons that is contained in a short GRB that has to be compared to the minimum number of counts ($\sim$ 100) that are needed to make an image with IBIS/ISGRI in order to confirm and localize the transients.

Another advantage of the instruments on board INTEGRAL is the broad energy range, from a few keV to the MeV region. Indeed, even if due to the limited field of view of JEM-X, only a few bursts have been detected down to a few keV, this region of the prompt GRB spectra is not well explored, and can be of great interest, see \cite{camarillobrillo13,minaev12}.

A further interesting result about the INTEGRAL GRB sample is the one provided by Foley et al. \cite{foley08}, that shows some degree of correlation (at $\sim$ 2.5 $\sigma$ level) of the GRBs presenting long ($>$ 1 s) time lags  between the 25--50 keV and the 50--300 keV energy bands, and the Super-Galactic Plane. This correlation indicates the possible "local" origin of these events, and hence suggests the idea of the existence of an intrinsically faint population of GRBs.

\section{GRB 031203: a remarkable GRB}
GRB 031203 is for sure one of the most remarkable GRBs detected with INTEGRAL for many reasons. First of all, because it is one of the few GRBs that are firmly spectroscopically associated with type Ib/c Supernovae (GRB980425 with SN 1998bw, GRB 030329/SN 2003dh, GRB 031203/SN 2003lw, GRB 060218/SN 2006aj, GRB100316D/SN 2010bh, GRB120422A/SN 2012bz). The Supernova associated to GRB 301203, located at z=0.01055, is very luminous reaching a rest-frame magnitude of M$_{V}$=-19.75 \cite{malesani04}, while the analysis of the IBIS data indicates that the GRB is, at variance, quite faint (<6$\times$10$^{49}$ erg<E$_{ISO}$<1.4$\times$10$^{50}$ erg) with a moderately high peak energy (E$_{Peak}$ > 190 keV) \cite{sazonov04}, making it a clear outlier of the E$_{Peak}$-E$_{ISO}$ relation that holds for most GRBs \cite{amati07}.

\begin{figure}
   \centering
   \includegraphics[width=0.4\textwidth]{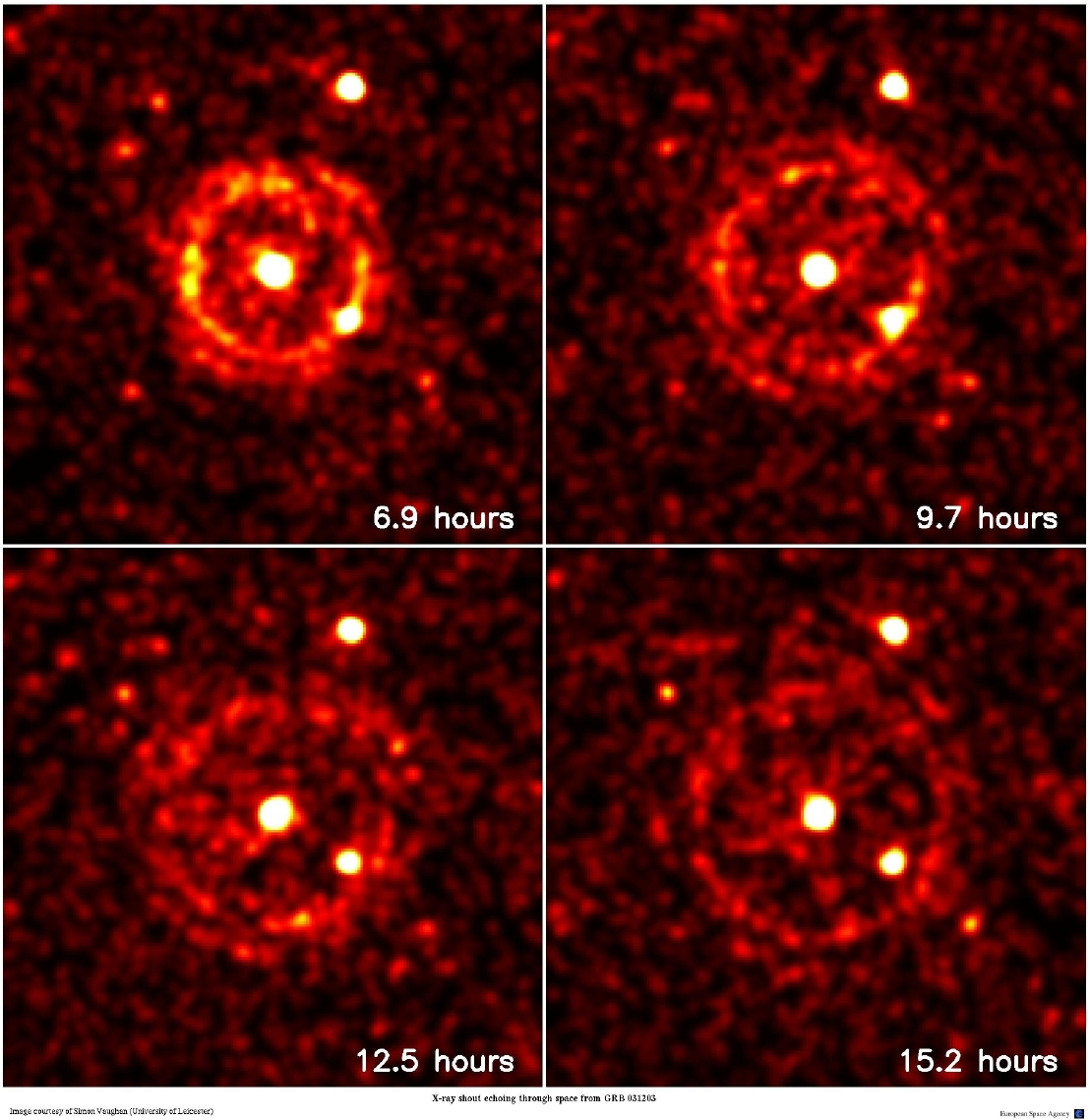}
   \includegraphics[width=0.5\textwidth]{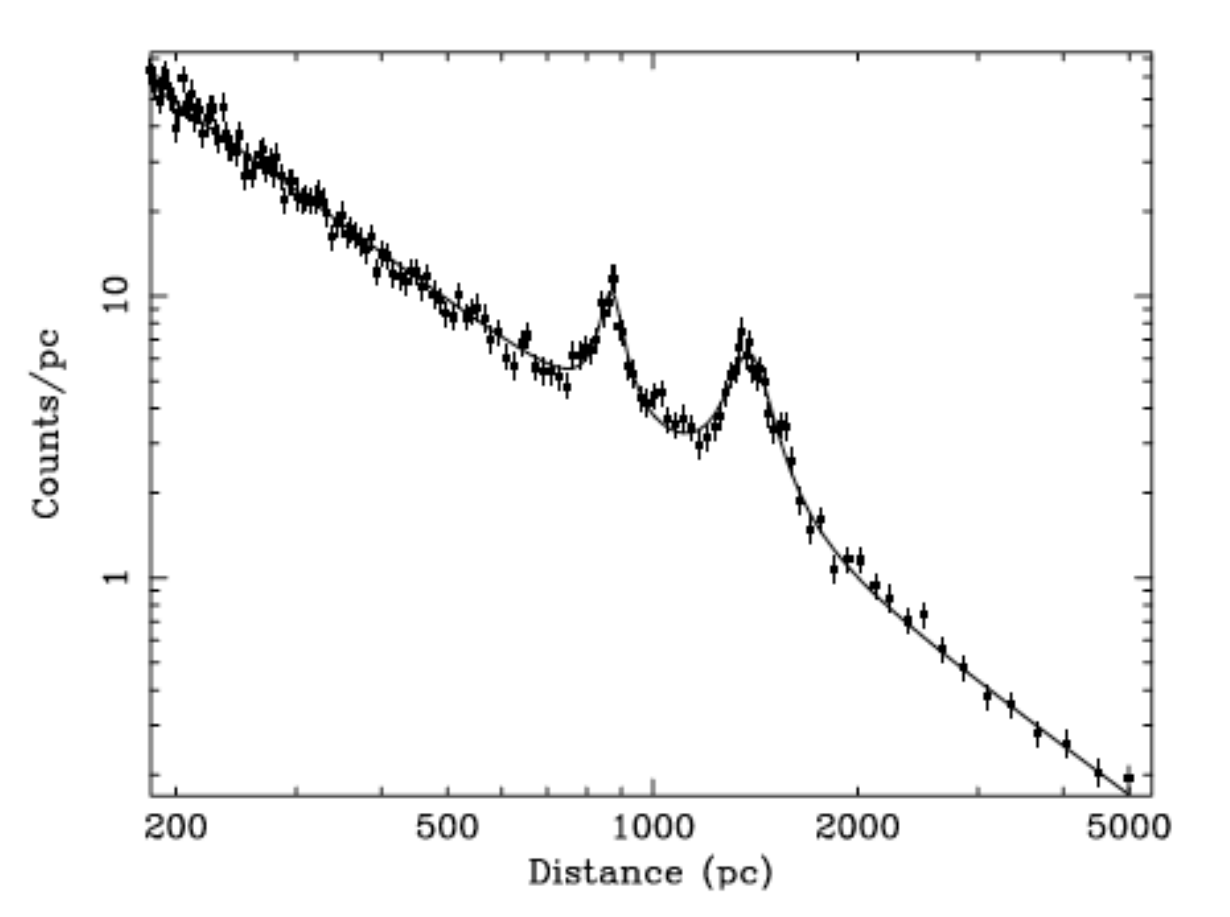}
   \caption{\textit{Left:} XMM images of the evolving dust scattering halo associated to GRB 031203 \cite{vaughan04}. \textit{Right:} Modelling of the two scattering rings constraining their distance \cite{tiengo06}.}
   \label{fig:halo}
   \end{figure}

The second reason that makes of GRB 031203 a very interesting burst is that, thanks to a quick follow-up observation performed by the XMM-Newton X-ray observatory, two evolving dust scattering halos have been discovered for the first time \cite{vaughan04}, see Fig. \ref{fig:halo}. Those haloes, or scattering rings, are due to the scattering of X-rays associated to the GRB prompt emission by dust clouds located in our own Galaxy. Indeed through an accurate modelling of the X-ray data Tiengo \& Mereghetti \cite{tiengo06} have been able to precisely locate those dust clouds showing that GRBs can be used to investigate the structure of the Galaxy, see Fig. \ref{fig:halo}. 

But the analysis of the scattered flux can be used also to derive the prompt X-ray flux of GRB 031203. As shown by Watson et al. \cite{watson04}, the prompt X-ray flux is not compatible in terms of fluence and spectrum with the prompt $\gamma$-ray spectrum, as measured with INTEGRAL. Some arguments have been put forward to explain the presence of this bright X-ray counterpart, undetected in the IBIS data, like a GRB seen off-axis \cite{ramirez-ruiz05}, the presence of a scattering cloud close to the GRB site, or a strong hard-to-soft spectral evolution, implying a long living X-ray component \cite{ghisellini06}. But despite these explanations, GRB 031203 remains a peculiar GRB with many uncommon characteristics. 

\section{GRB 041219A: the record holder}
GRB 041219A is another INTEGRAL GRB for which many interesting results could be derived. This is mainly due to the fact that this GRB is one on the longest, and brightest ever detected. Indeed it lasts about 500 s, and, with a 20-200 keV fluence reaching 2.5$\times$10$^{-4}$ erg cm$^{-2}$, it is the brightest GRB in the INTEGRAL sample, but ranking also in the top 1\% of the BATSE bursts.

\subsection{Broad band modelling}

\begin{figure}
   \centering
   \includegraphics[width=0.55\textwidth]{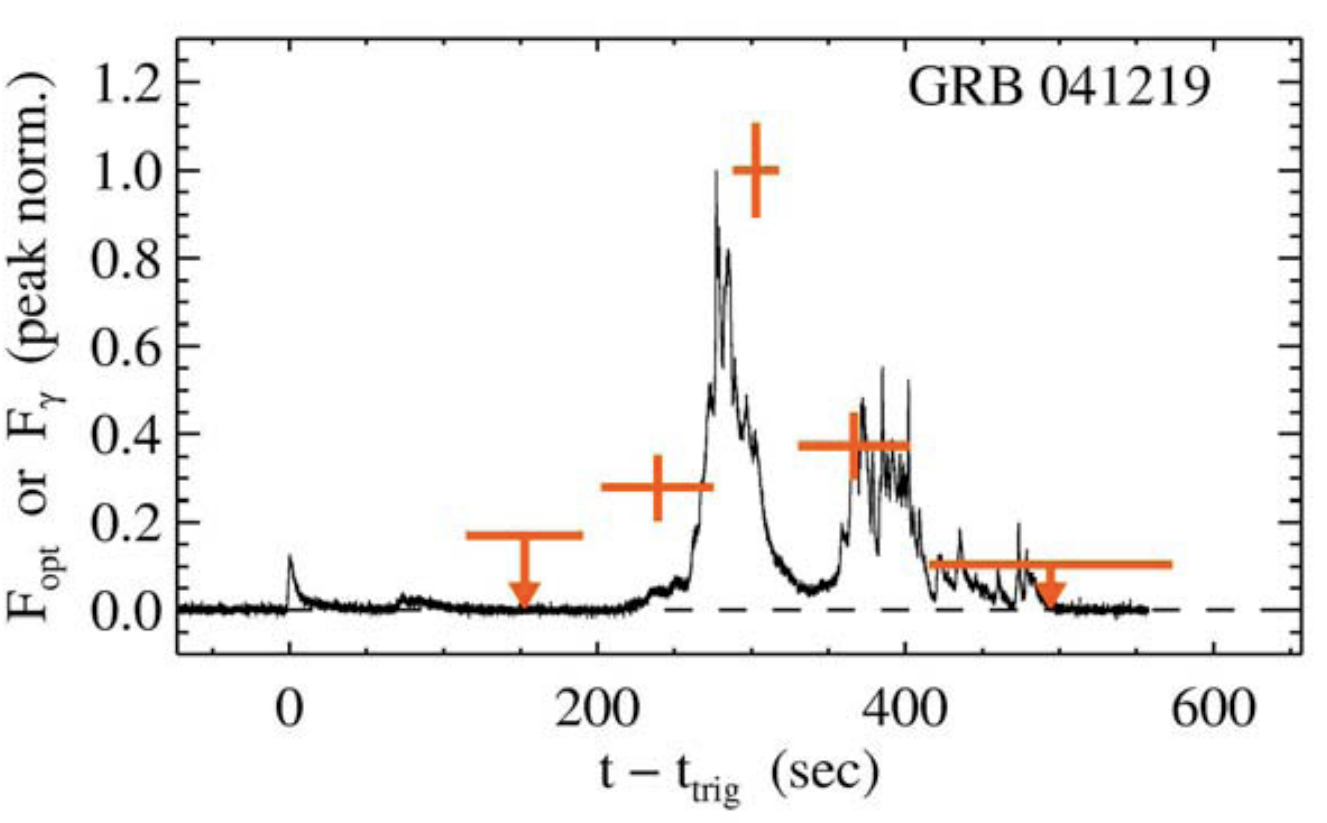}
      \includegraphics[width=0.3\textwidth]{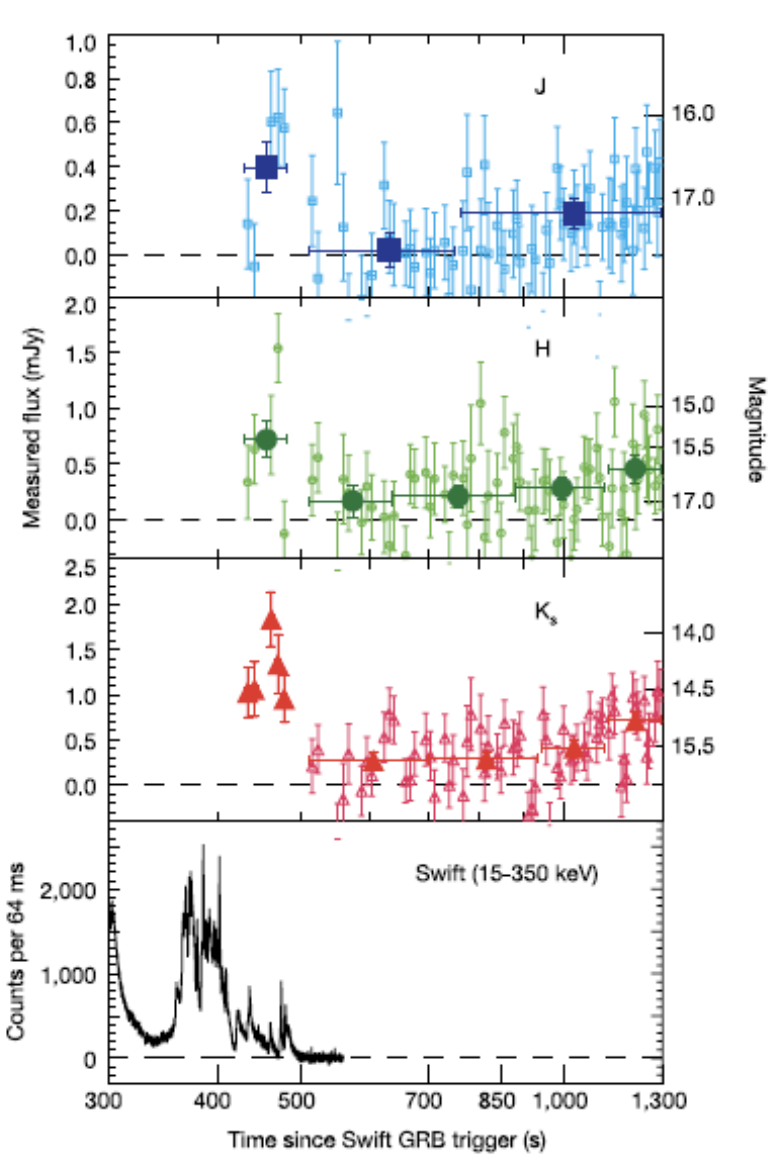}

   \caption{\textit{Left:} $\gamma$ vs. optical light curve of GRB 041219A. From \cite{vestrand05}. \textit{Right:} NIR multi filter light curve of the final part of GRB 041219A. From \cite{blake05}.}
   \label{fig:041219lc}
   \end{figure}

Due to its long duration and to the fact that an IBAS alert was issued on a precursor taking place about 250 before the main events, robotic telescopes had the time to point the GRB prompt emission. This makes of GRB 041219A one of the few GRBs for which a simultaneous covering in the $\gamma$-ray, optical and NIR bands is available. Vestrand et al. \cite{vestrand05} report that the level of optical flux seems to be correlated to the gamma one, see Fig. \ref{fig:041219lc}, while Blake et al. \cite{blake05} have been able to collect some multi band NIR data on the final part of the burst. These multi wavelength data, spanning over six decades in energy, are very important because they can help to constrain the nature of the prompt emission of GRBs, a still unsettled question.

For instance G\"otz at al. \cite{gotz11} took advantage from this very rich dataset to try to physically model the prompt emission. They found that in the classical framework of internal shocks, assumed for GRB modelling, where the radiation observed is explained by synchrotron emission produced by relativistic electrons
accelerated by the propagating shock waves (e.g. \cite{daigne98}), the broad band emission can be explained only if the deceleration of the ejecta by the external medium, and more generally the role of the reverse shock, is taken into account. But this modelling holds only for the optical to $\gamma$-ray broad band spectrum. Indeed, once the NIR data are included in the fit, they do not agree neither in terms of shape nor in flux with the model. Some explications for the different origin of the NIR component have been proposed, like the rising of the external shock or late internal collisions due to late time activity of the central engine, but both scenarios do not work properly. 

\begin{figure}
   \centering
   \includegraphics[width=0.47\textwidth]{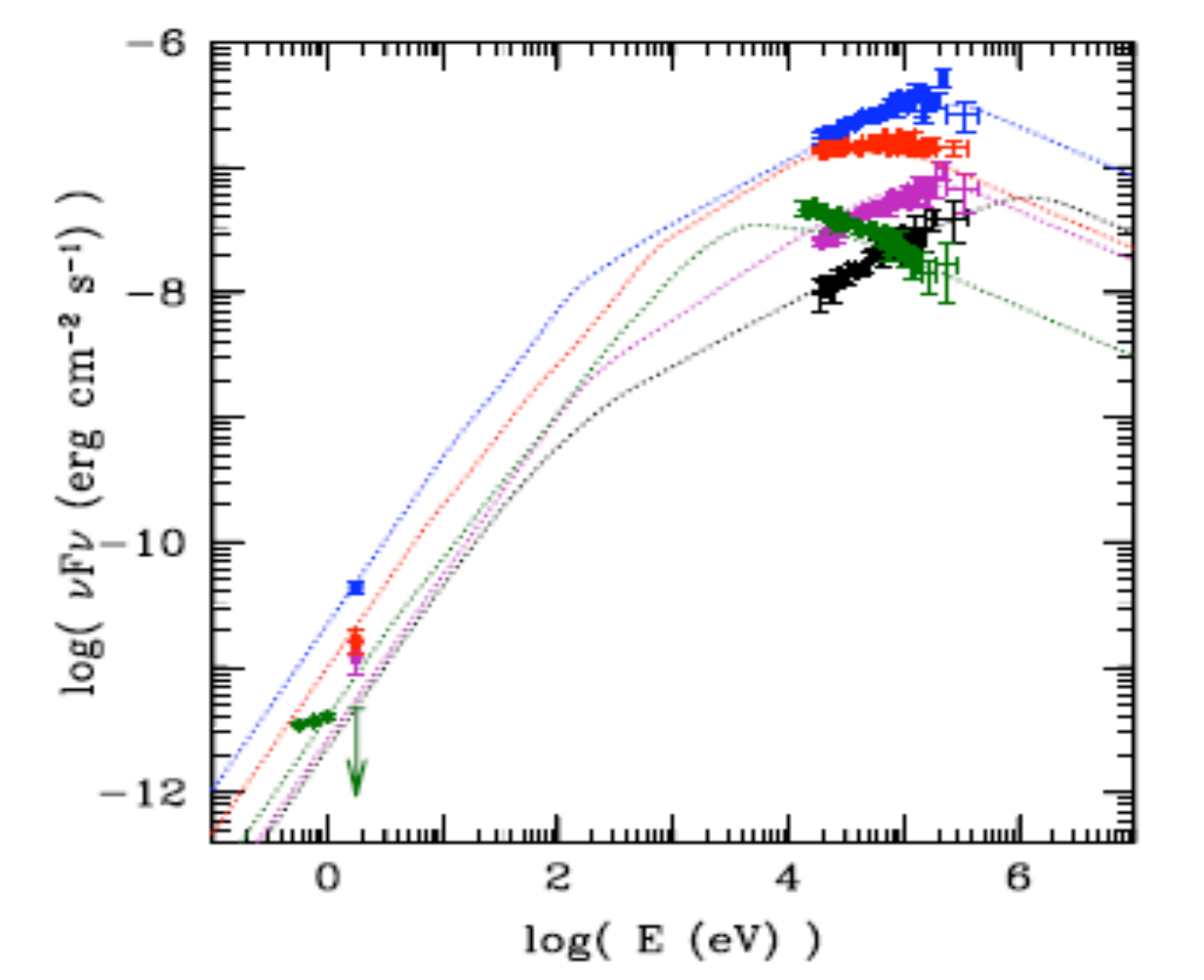}
      \includegraphics[width=0.52\textwidth]{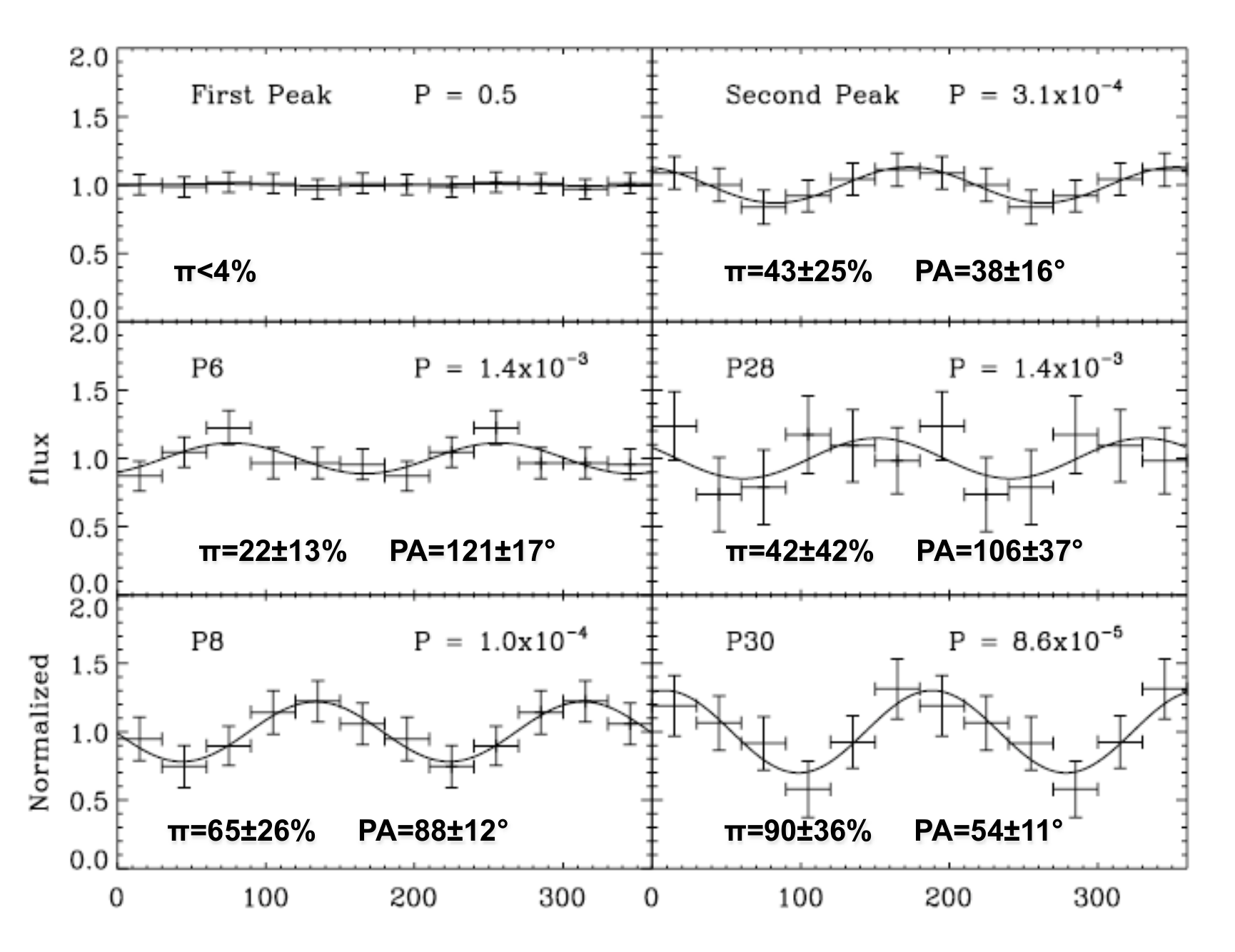}

   \caption{\textit{Left:} Multi wavelength modelling of GRB 041219A. The different colours represent the different time intervals on which the analysis was possible. The green data represent the time interval for which the NIR data were available. From \cite{gotz11}. \textit{Right:}  From \cite{gotz09}}
   \label{fig:041219}
   \end{figure}

\subsection{Polarization}

Despite the richness of the GRB 041219A dataset the nature of its prompt emission is not completely clarified. Another tool that can be used to investigate the nature and the geometry of the prompt emission of GRBs is polarimetry: the two main instruments on board INTEGRAL, IBIS and SPI, have some polarimetric capabilities, provided that the sources that are examined are bright enough. Due to large number of detected counts several authors have tried to measure the polarization signals from this GRB \cite{kalemci07,mcglynn07,gotz09}. In particular McGlynn et al.\cite{mcglynn07} and G\"otz et al. \cite{gotz09} reported some positive detection of a high level of polarization using SPI and IBIS respectively. The IBIS results are of particular interest, since they show that the level of polarization and the polarization angle vary with time along the burst, see Fig. \ref{fig:041219}. Despite the fact that this single measurement can not give a definite answer to the process producing the polarized signal (see discussion in \cite{gotz09}), and that more measurements on different GRBs are needed to clarify the global picture, the variability points towards an interpretation where synchrotron radiation is emitted from shock accelerated electrons in a relativistic jet with a magnetic field transverse to the jet expansion. The coherence of the magnetic field geometry does not need to hold over the entire jet but only over a small portion of it since, due to relativistic effects, the observer can see only a region of the jet whose angular size comparable to 1/$\Gamma$, $\Gamma$ being the Lorentz factor of the relativistic outflow. If the radiating electrons are accelerated in internal shocks, then the Lorentz factor is necessarily varying in the outflow, which can explain the variability of the polarization from one pulse to the other.

High level of polarization, sometimes variable, has been reported also for other GRBs observed with the dedicated Japanese GAP experiment \cite{yonetoku11,yonetoku12}, and a complete analysis of the INTEGRAL GRB sample is on-going and a few good candidates have already been identified. Increasing the sample of GRBs with measured polarization will indeed help to discriminate among competing models, that can not be completely ruled out at the time -- like 
"Compton drag", fragmented fireballs, purely electromagnetic flows, etc. --, confirming that polarization is a very important complementary way to investigate the nature of the radiation produced by GRBs. 

\subsection{Constraints on the Lorentz Invariance Violation}
On general grounds one expects that the two fundamental theories of contemporary physics, the theory of General Relativity and the quantum theory in the form of the Standard Model of particle physics, can be unified at the Planck energy scale. This unification requires to quantize gravity, which leads to very fundamental difficulties: one of these is the possibility of Lorentz Invariance Violation (LIV).

A possible experimental test of LIV is testing the helicity dependence of the propagation velocity of photons, see Eq. \ref{eq:dispersion1}:

\begin{equation}
\omega^{2}=k^{2}\pm\frac{2\xi k^{3}}{M_{Pl}}\equiv\omega^{2}_{\pm}
\label{eq:dispersion1}
\end{equation}

\begin{equation}
\omega_{\pm}=\vert p \vert \sqrt{1\pm\frac{2\xi k}{M_{Pl}}}\approx\vert k\vert(1\pm\frac{\xi k }{M_{Pl}})
\label{eq:dispersion2}
\end{equation} 

\noindent where $E=\hbar\omega$, $p=\hbar k$, and $\xi$ gives the order of magnitude of the effect. In other words, if a polarized signal is measured from a distant source, some quantum-gravity theories predict that the polarization plane should rotate by a quantity $\Delta\theta$ while the electromagnetic wave propagates through space, and this as a function of the energy of the photons. This is illustrated in Eq. \ref{eq:rotation}, where $d$ is the distance of the source.

\begin{equation}
\Delta\theta(p)=\frac{\omega_{+}(k)-\omega_{-}(k)}{2}d\approx\xi\frac{k^{2}d}{2M_{Pl}}
\label{eq:rotation}
\end{equation}

This would imply that a polarized signal produced by a given source would vanish if observed on a broad band, since the differential rotation acting on the polarization angle as a function of energy would in the end add opposite oriented polarization vectors.
But being this effect very tiny, since it is inversely proportional to the the Planck Mass ($M_{Pl}\sim$2.4$\times$10$^{18}$ GeV), the observed source needs to be at cosmological distances. So, the simple fact to detect the polarization signal from a distant source, can put a limit to such a possible violation. This experiment has been performed by Maccione et al. \cite{maccione08} using the SPI measure of the polarization from the Crab nebula, and constraining $\xi$<2$\times$10$^{-9}$.

\begin{figure}
   \centering
   \includegraphics[width=0.45\textwidth]{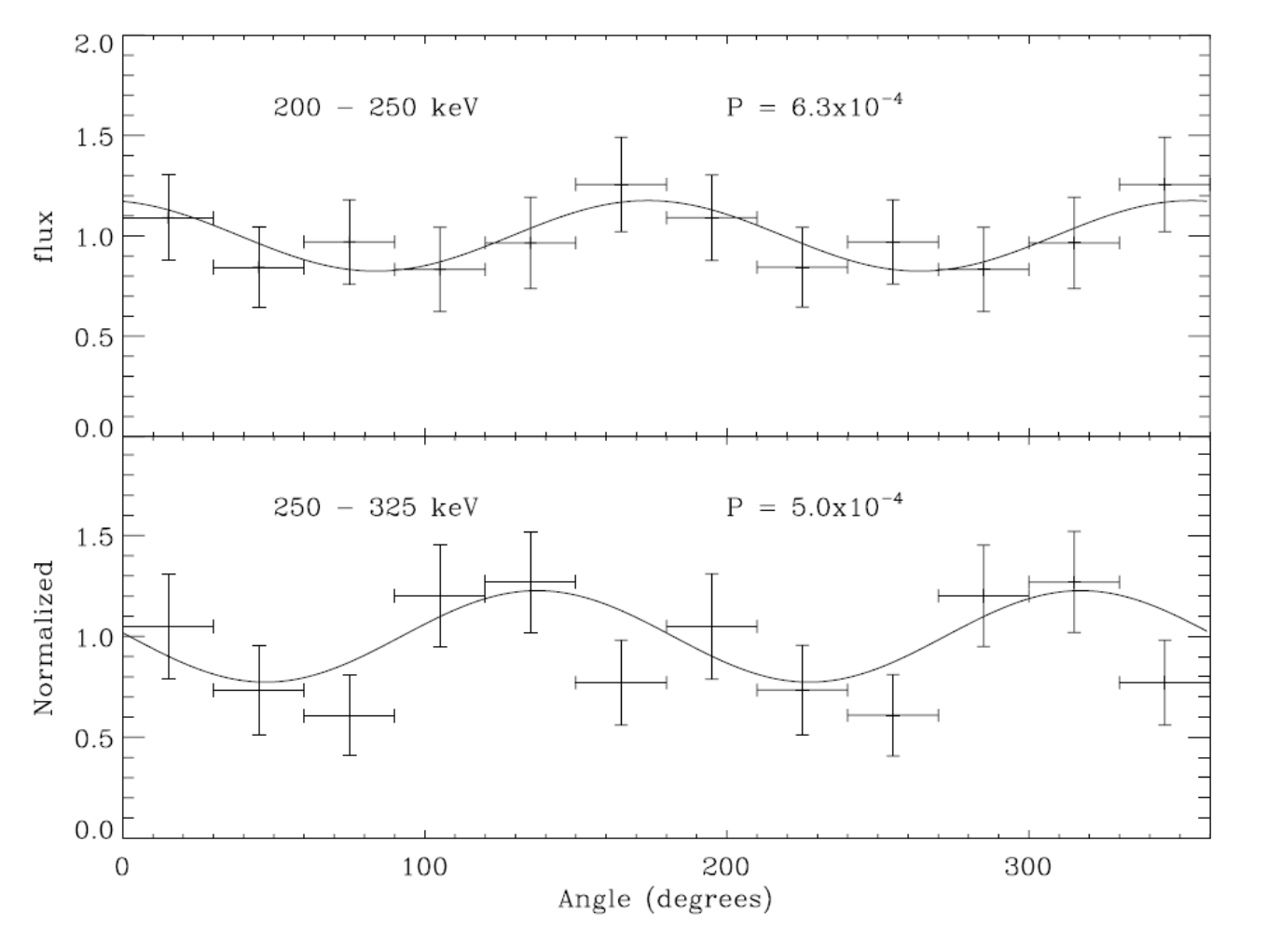}
      \includegraphics[width=0.5\textwidth]{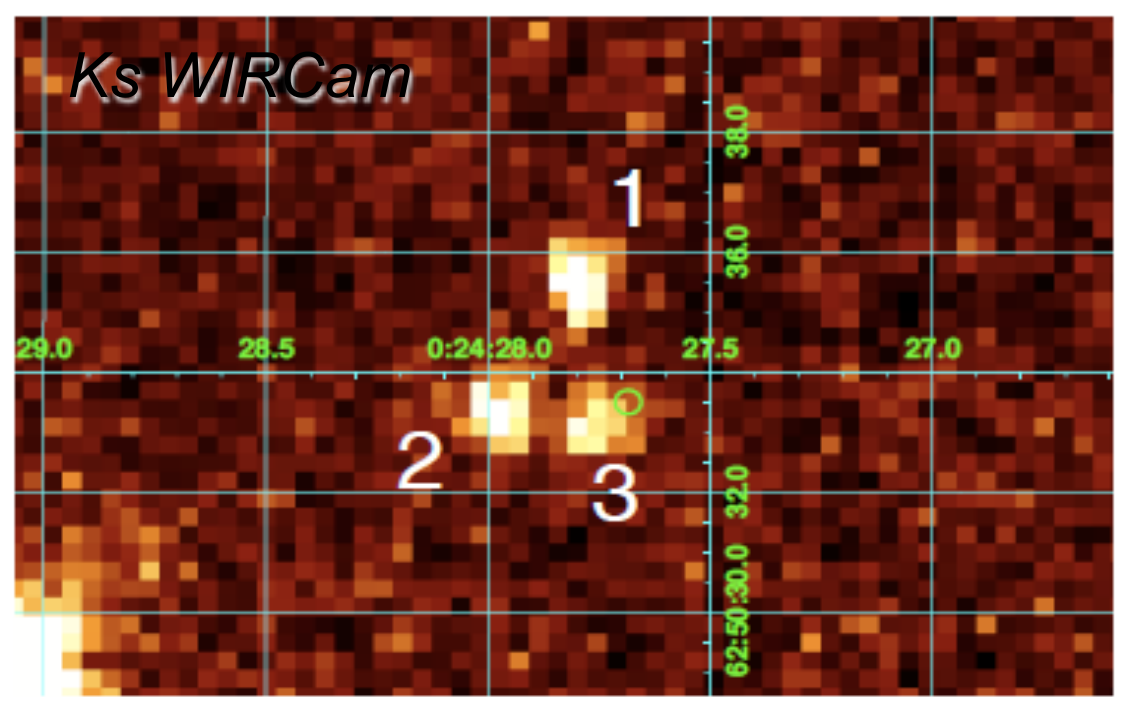}
   \caption{\textit{Left:} Energy resolved polarization measurements of GRB 041219A. From \cite{laurent11}. \textit{Right:} CFHT WIRCam $K_{S}$ image of the field of GRB 041219A. The GRB host has been identified as object 3. The circle represent the error box derived from the optical afterglow. From \cite{gotz11}.}
   \label{fig:liv}
   \end{figure}

Being Gamma-Ray Bursts much more distant than the Crab nebula, they are very good candidates to further improve those limits by several orders of magnitude. This has been performed by Laurent et al. \cite{laurent11}, taking advantage from the polarization measurements obtained with IBIS on GRB 041219A in different energy bands (200--250 keV, 250--325 keV), and from the measurement of distance of the source (z=0.31$^{+0.54}_{-0.26}$, equivalent to a distance interval of [0.222--5.406] Gpc) derived from follow-up observations of the GRB host galaxy by \cite{gotz11}, see Fig. \ref{fig:liv}. They were able to greatly improve the limit on the LIV to $\xi$<1.1$\times$10$^{-14}$.

\section{Conclusions and lessons learned}
INTEGRAL GRB studies have risen more questions than provided answers, but this is probably true also for all recent missions with GRB capabilities, like Swift or Fermi. Nevertheless INTEGRAL has shown the potential of polarization studies for GRB science (and fundamental physics!), pointing out the necessity of this kind of experiment in the future. In addition the INTEGRAL experience has pointed out that general purpose X-ray and $\gamma$-ray observatories can often be used for GRB science at a relatively low additional cost, providing interesting results for a broad community. This implies that a wide field of view, possibly "all-sky", X-ray monitoring experiment is mandatory for the near future, not strictly for GRB science, but also because it provides the necessary inputs for other wavelengths/astrophysical domains and upcoming observatories (GW detectors, neutrino, large radio arrays, etc.).


\begin{thebibliography}{99}
\bibitem{integral}{Winkler} C. {et~al.}, 2003, A\&A, 411, L1
\bibitem{atteia13}{Att\'eia} J.L., 2013, these proceedings
\bibitem{ibas}{Mereghetti} S. et al., 2003, A\&A,  411, L291
\bibitem{mereghetti13}{Mereghetti} S., et al. 2013, these proceedings
\bibitem{ibis}{Ubertini} P. {et~al.}, 2003, A\&A, 411, L131
\bibitem{isgri}{Lebrun} F. {et~al.}, 2003, A\&A, 411, L141
\bibitem{isdc}Courvoisier T.J.-L. et al., 2003, A\&A,  411, L53
\bibitem{spi}Vedrenne G. et al., 2003, A\&A,  411, L63 
\bibitem{bosnjak13}Bo\v snjak Z. et al., 2013, these proceedings 
\bibitem{camarillobrillo13}Martin Carrillo A. et al., 2013, these proceedings 
\bibitem{minaev12}Minaev, P. Yu. et al., 2012, AstL, 38, 613
\bibitem{foley08}Foley S. et al., 2008, A\&A,  484, 143 
\bibitem{malesani04}Malesani D. et al., 2004, ApJ, 609, L5 
\bibitem{sazonov04}Sazonov S. Yu. et al., 2004, Nature, 421, L21
\bibitem{amati07}Amati L., 2007, MNRAS, 372, 233
\bibitem{vaughan04}Vaughan S. et al., 2004, ApJ, 603, L5
\bibitem{tiengo06}Tiengo A. \& Mereghetti S., 2006, A\&A, 449, 203
\bibitem{watson04}Watson D. et al, 2004, ApJ, 605, L101
\bibitem{ramirez-ruiz05}Ramirez-Ruiz E. et al, 2005, ApJ, 625, L91
\bibitem{ghisellini06}Ghisellini G. et al, 2006, MNRAS, 372, 1699
\bibitem{vestrand05}Vestrand, W.T. et al., 2005, Nature, 435, 178 
\bibitem{blake05}Blake C.H. et al., 2005, Nature, 435, 181 
\bibitem{gotz11}G\"otz D. et al., 2011, MNRAS, 413, 2173 
\bibitem{daigne98}Daigne F. \& Mochkowitch R., 1998, MNRAS, 296, 275 
\bibitem{kalemci07}Kalemci E. et al., 2007, ApJS, 169, 75 
\bibitem{mcglynn07}McGlynn S., 2007, A\&A, 466, 895
\bibitem{gotz09}G\"otz D. et al., 2009, 695, L208 
\bibitem{yonetoku11}Yonetoku, D. et al., 2011, ApJ, 743, L30
\bibitem{yonetoku12}Yonetoku, D. et al., 2012, ApJ, 758, L1
\bibitem{maccione08}Maccione L. et al., 2008, Physical Review D, 78, 103003
\bibitem{laurent11}Laurent P. et al., 2011, Phys. Rev. D, 83, 121301(R) 

\end{thebibliography}
\end{document}